# ATM: a Logic for Quantitative Security Properties on Attack Trees

Stefano M. Nicoletti[1*], Milan Lopuhaä-Zwakenberg[1], E. Moritz Hahn[1], Mariëlle Stoelinga[1,2]

[1*]Formal Methods and Tools, University of Twente, Enschede, the Netherlands.
[2]Department of Software Science, Radboud University, Nijmegen, the Netherlands.

*Corresponding author(s). E-mail(s): s.m.nicoletti@utwente.nl;
Contributing authors: m.a.lopuhaa@utwente.nl; e.m.hahn@utwente.nl;
m.i.a.stoelinga@utwente.nl;

**Abstract**

Critical infrastructure systems — for which high reliability and availability are paramount — must operate securely. Attack trees (ATs) are hierarchical diagrams that offer a flexible modelling language used to assess how systems can be attacked. ATs are widely employed both in industry and academia but — in spite of their popularity — little work has been done to give practitioners instruments to formulate queries on ATs in an understandable yet powerful way. In this paper we fill this gap by presenting **ATM**, a logic to express quantitative security properties on ATs. **ATM** allows for the specification of properties involved with *security metrics* that include "cost", "probability" and "skill" and permits the formulation of insightful what-if scenarios. To showcase its potential, we apply **ATM** both to the case study of a CubeSAT and to a larger model, constructed from the real-life cyberespionage campaign *Operation Dream Job*, as recorded by the MITRE ATT&CK Database. We showcase property specification on the corresponding attack trees and propel usability of **ATM** by presenting **LangATM** – a domain specific language for our logic. Finally, we present theory and algorithms — based on binary decision diagrams — to check properties and compute metrics of **ATM**-formulae.

## 1 Introduction

Critical infrastructure systems — for which high reliability and availability are paramount — must operate securely. Attack trees (ATs) [1] are a flexible modelling language used to assess how systems can be attacked. They operate by decomposing the attacker's goal into intermediate elements and basic attack steps that a malicious actor can take to reach said objective. ATs are widely employed both in industry and academia but — in spite of their popularity — little work has been done to give practitioners instruments to formulate queries on ATs in an understandable yet powerful way. In this paper, we fill this gap by presenting a logic to express quantitative Metrics on ATs (ATM). ATM is a powerful language able to formulate structural queries on ATs that consider quantitative security properties, or *security metrics*, such as "cost" of an attack, "probability" of getting attacked and "skill" of a malicious actor. The ability to formulate these queries is essential to provide practitioners with an instrument to analyse what-if scenarios and to propel a more quantitatively-informed decision making process.

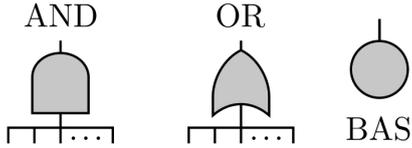

**Figure 1**: Nodes in an attack tree.

**Attack trees.** Attack trees (ATs) are hierarchical diagrams that represent various ways in which a system can be compromised [1, 2]. Due to their popularity, ATs are referred to by many system engineering frameworks, e.g. *UMLsec* [3] and *SysMLsec* [4, 5], and are supported by industrial tools such as Isograph's *AttackTree* [6]. The root — or *top level event* (TLE) — of an AT represents the attacker's goal, and the leaves represent *basic attack steps* (BASes): actions of the attacker that can no longer be refined. Intermediate nodes are labeled with gates (see Fig. 1) that determine how basic actions of the attacker can propagate to reach higher-complexity elements in the attack. ATs that do not capture dynamic behaviours present only OR and AND gates — we call these *static attack trees* (SATs) — but many extensions exist to model more elaborate attacks. To build a solid and modular foundation for our framework, this paper focuses on SATs. It is important to note that — despite their name — ATs can be *directed acyclic graphs* (DAGs), i.e., graphs in which a node may have multiple parents. Them being DAGs or tree-structured has consequences on computations [2].

**Example 1.** *Consider the AT in Fig. 2 (excerpt from Fig. 6). This AT represents different attacks to get access to the ground station database of a CubeSAT as admin. The TLE of this sub-tree is represented by the ADA AND-gate. For the attacker to reach ADA, they have to both gain access — the GA AND-gate — and escalate privileges — the EP OR-gate. Each of these gates is then refined by BASes: to gain access, an attacker must perform information gathering and a successful phishing attack — the IGP BAS— and login to the ground station database using phished credentials —the LDG BAS. In addition, they have to either leverage misconfigurations — the LM BAS— or exploit vulnerabilities — the EV BAS— to escalate privileges. Note that IGP is represented here as a BAS but is further refined as an additional sub-tree in Fig. 6.*

**Metrics on attack trees.** ATs are often studied via *quantitative analysis*, during which they are assigned a wide range of security metrics [2]. Such metrics are key performance indicators that formalize how well a system performs in terms of security and are essential when comparing alternatives or making trade-offs. Typical examples of such metrics are the minimal time [7–10], minimal cost [11], or maximal probability [12] of a successful attack (see Table 1 for more examples).

## 1.1 Our approach

**A logic for Attack Trees Metrics (ATM).** To perform quantitatively informed decision making w.r.t. security of systems, practitioners need the ability to analyse their models in a meaningful and thorough way. As such, they must be able to formulate *meaningful queries* and meaningful *what-if scenarios*. To cater for this need, this paper presents ATM, a logic for general Metrics on Attack Trees. ATM is a flexible language used to specify properties that take metrics such as "cost", "skill" and "probability" into account directly on ATs. ATM is structured on four layers: these allow practitioners a) to reason about *successful/unsuccessful* attacks; b) to check whether metrics, such as the cost, are bounded by a given value on single attacks; c) to compute metrics for a class of attacks and d) to perform quantification.

**Attack trees in practice.** To offer a concrete example, we utilise ATM to specify some properties on the AT model of a CubeSAT [13, 14] from the literature [15] (see Fig. 6). This model exemplifies the effect of a security threat for the availability of a system by showcasing three ways in which a malicious actor could attack a CubeSAT: performing a denial of service attack, tampering with data on the database of the ground station or killing radio communications on the satellite. Our logic can be used to specify properties on the corresponding AT and to check whether the system under examination exhibits desired characteristics. Is it necessary to leverage misconfigurations to perform a successful attack on CubeSAT's communications? Is there an attack that ensures successful access to the ground station database while keeping the cost under a certain threshold? Is there an attack that ensures data tampering without exploiting vulnerabilities in the ground station system? To further showcase real-life usefulness of our approach, we construct an AT model from a real-life cyberespionage campaign on the aerospace sector – *Operation Dream Job* – as recorded by the MITRE ATT&CK Database[★], a globally-accessible knowledge base of adversary tactics and techniques based on real-world observations. Can an attacker always establish presence in the network only by guaranteeing execution of all (sub)techniques for the resource development and discovery tactics? Is it sufficient to perform defense evasion techniques in order to achieve both privilege escalation and persistence? Is there a successful attack for the overall system that does not require to evade sandboxes? These are some of the properties that one could specify and check in the framework we present.

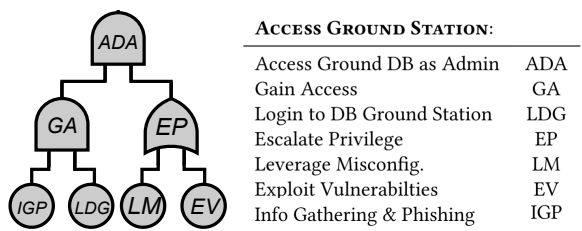

| Access Ground Station: | |
|---|---|
| Access Ground DB as Admin | ADA |
| Gain Access | GA |
| Login to DB Ground Station | LDG |
| Escalate Privilege | EP |
| Leverage Misconfig. | LM |
| Exploit Vulnerabilties | EV |
| Info Gathering & Phishing | IGP |

**Figure 2**: AT modelling access to ground station DB of a CubeSAT (excerpt from Fig. 6).

**Model checking algorithms.** In addition, we present model checking algorithms to check properties specified with ATM and to efficiently compute metrics that appear in these properties. In particular, we provide algorithms to a) check whether an AT and an attack satisfy a formula; b) compute all attacks that satisfy an AT and a formula; c) check whether the metric of a formula is bounded by a user-specified threshold; d) compute the metric value of formulae and e) check whether a quantified ATM-formula holds true. Building on previous work in the field [2, 16, 17], all these algorithms are based on construction and manipulation of binary decision diagrams (BDDs). This translation to BDDs constitutes a formal ground to address algorithmic procedures while integrating novel work presented in this paper with previously introduced frameworks.

This paper extends previous work [18]:

**Contributions of [18]:**
1. We develop ATM, a logic to reason about general metrics on ATs. ATM allows for the specification of metrics properties that include "cost", "probability" and "skill" and for the formulation of insightful what-if scenarios.
2. We showcase ATM by applying it to the case study of a CubeSAT and by exemplifying properties specification.
3. We propose novel algorithms based on BDDs to perform model checking and to compute metrics on properties specified using ATM.

**Additional contributions of this version:**
1. To showcase real-life usefulness of our approach, we apply ATM to a newly developed attack tree instance for the Operation Dream Job cyberespionage campaign, constructed from MITRE ATT&CK.
2. To propel usability, we define LangATM, a domain specific language for ATM.
3. Furthermore, we considerably extend our analysis of related work, taking other approaches for metrics computation on ATs into account.

### 1.2 Related work

Numerous logics describe properties of state transition systems, such as labelled transition systems (LTSs) and Markov models, e.g., CTL [19], LTL [20], and their variants for Markov models, PCTL [21] and PLTL [22]. State-transition systems are usually not written by hand, but are the result of the semantics of high-level description mechanisms, such as AADL [23], the hardware description language VHDL [24] or model description languages such as JANI [25] or PRISM [26]. Consequently, these logics are not used to reason about the structure of such models (e.g. the placement of circuit elements in a VHDL model or the structure of modules in a PRISM model), but on the temporal behaviour of the underlying state-transition system. Similarly the majority of related work [27–30] on model checking on *fault*

---

[★] https://attack.mitre.org/

*trees* (FTs) — the safety counterpart of ATs— exhibits significant differences: these works perform model checking by referring to states in the underlying stochastic models, and properties are formulated in terms of these stochastic logics, not in terms of events in the given FT. In [31], the author provides a formulation of *Pandora*, a logic for the qualitative analysis of temporal FTs. In [32] the authors investigate how fault tree analysis (FTA) results can be linked to software safety requirements by proposing the same system model for both. They introduce a duration calculus based on discrete time interval logic (ITL) [33] to give FTs formal semantics. In [16, 17] we present BFL — a logic on FTs that reasons about them in Boolean terms — and PFL — its probabilistic counterpart. Our work is aligned in intentions to the latter two, as we develop a logic directly on ATs. As for FTs, our queries allow not only for qualitative analysis, but also quantitative analysis via metric queries; however, a large difference between FTs and ATs is that in security analysis one does not only consider the single probability metric, but a wide range of security metrics such as required attacker cost [34, 35], attack time [7, 9], attacker skill [2, 7], etc. While work exists that only considers one of these metrics, most work on AT analysis follows Mauw & Oostdijk [36] in phrasing AT metrics in terms of semirings, an algebraic structure consisting of a set and two operators, corresponding to AND- and OR-gates. As a framework, it is general enough to capture most metrics of interest, while having enough mathematical structure to allow for efficient computation.

Unfortunately, different works have slightly different definitions of how the structure of an AT and a semiring together define the AT's security value. An overview of these definitions, along with a category-based generalization that encompasses all of these, is given in [37]. We can roughly distinguish three approaches: On one hand, the metric value can be defined bottom-up, using the operators corresponding to the gate types [36]. By definition this is very quickly computed, however, it assumes that a BAS has to be performed multiple times, once for every parent it activates. Thus this approach cannot model situations in which attacker actions have multiple independent consequences, limiting the expressivity of this approach.

The second approach is to first bottom-up define a set of considered attacks, and then use the semiring structure to calculate the metric value over this set of attacks [38]. This approach detects when a BAS is used multiple times, and counts each BAS only once in the metric computation. Unfortunately, this makes calculation also more complicated: the existing approach works by splitting the extra appearances of BAS in so-called optional and necessary clones and accounting for this in the calculation; this process is exponential in the number of shared BASes.

A downside of the second approach is that the bottom-up defined set of attacks can contain some non-optimal attacks, skewing the metric computation. To address this fact, the third approach considers a set of *minimal attacks* defined from the semantics, and then calculates the metric on those [2]. Metrics defined in this way can be computed by translating the AT to a BDD, which is worst-case exponential but quite fast in practice. In this paper, we will follow this approach as it is expressive, and its BDD-based approach can be extended to more complicated logic queries.

Besides semiring-based definitions, metrics can also be defined/calculated by translating the AT into an automata model [7, 39, 40] and using tools such as UPPAAL [41] for metric calculation; this has been applied to metrics such as attack time, attack cost, and attack probability. The downside of this approach is that it does not generalize to arbitrary semirings. Furthermore, since a formal definition of the metric is lacking, the metric as calculated in this way can show undesired behaviour [42].

Several extensions of the AT framework exist. Some of the most prominent are: dynamic attack trees, in which the order of the BASes is expressed through sequential AND-gates [7, 42, 43]; attack-fault trees, combining safety and security analyses in a single framework [39, 40]; and attack-defense trees, which incorporate countermeasures [38, 44, 45]. Quantitative analysis on these models is generally done by extending one of the four methods above; for dynamic attack trees and attack-defense trees see [37].

**Structure of the paper.** Sec. 2 covers background on ATs, Sec. 3 presents syntax and semantics for ATM, Sec. 4 showcases an application of ATM to a CubeSAT AT, Sec. 5 further applies

ATM to a more complex AT representing a real-life cyberespionage campaign from the MITRE ATT&CK Database, Sec. 6 presents LangATM – a domain specific language for ATM, Sec. 7 presents model checking algorithms for ATM−formulae and Sec. 8 concludes the paper and discusses future work.

## 2 Attack Trees

**Definition 1.** *An* attack tree *(AT) $T$ is a tuple $(N, E, t)$ where $(N, E)$ is a rooted directed acyclic graph, and $t\colon N \to \{\mathtt{OR}, \mathtt{AND}, \mathtt{BAS}\}$ is a function such that for $v \in N$, it holds that $t(v) = \mathtt{BAS}$ if and only if $v$ is a leaf.*

Moreover, $ch\colon N \to \mathscr{P}(N)$ gives the set of *children* of a node and $T$ has a unique root, denoted $R_T$. The subindex $T$ is omitted if no ambiguity arises, e.g. an attack tree $T = (N, t, ch)$ defines a set $\mathrm{BAS} \subseteq N$ of basic attack steps. If $u \in ch(v)$ then $u$ is called a *child* of $v$, and $v$ is a *parent* of $u$. We let $v = \mathrm{AND}(v_1, \ldots, v_n)$ if $t(v) = \mathtt{AND}$ and $ch(v) = (v_1, \ldots, v_n)$, and analogously for OR, denoting $ch(v) = \{v_1, \ldots, v_n\}$. Furthermore, we denote the universe of ATs by $\mathscr{T}$ and call $T \in \mathscr{T}$ *tree-structured* if for any two nodes $u$ and $v$ none of their children is shared, else we say that $T$ is *DAG-structured*. If only AND- and OR-gates (or their derivatives) are present we say that $T$ is a *static attack tree* (SAT). In this paper we focus our attention on SATs and thus use the term ATs interchangeably to denote them. The semantics of a AT is defined by its successful attack scenarios, in turn given by its structure function. First, the notion of attack is defined:

**Definition 2.** *An* attack scenario, *or shortly an* attack, *of a static AT $T$ is a subset of its basic attack steps: $A \subseteq BAS_T$. We denote by $\mathscr{A}_T = 2^{BAS_T}$ the universe of attacks of $T$. We omit the subscript when there is no confusion.*

The structure function $f_T(v, A)$ indicates whether the attack $A \in \mathscr{A}$ succeeds at node $v \in N$ of $T$. For Booleans we adopt $\mathbb{B} = \{\mathtt{1}, \mathtt{0}\}$.

**Definition 3.** *The* structure function $f_T\colon N \times \mathscr{A} \to \mathbb{B}$ *of a static attack tree $T$ is given by:*

$$f_T(v, A) = \begin{cases} \mathtt{1} & \text{if } t(v) = \mathtt{OR} \ \text{and } \exists u \in ch(v).f_T(u, A) = \mathtt{1}, \\ \mathtt{1} & \text{if } t(v) = \mathtt{AND} \ \text{and } \forall u \in ch(v).f_T(u, A) = \mathtt{1}, \\ \mathtt{1} & \text{if } t(v) = \mathtt{BAS} \ \text{and } v \in A, \\ \mathtt{0} & \text{otherwise.} \end{cases}$$

An attack $A$ is said to *reach* a node $v$ if $f_T(v, A) = \mathtt{1}$, i.e. it makes $v$ succeed. If no proper subset of $A$ reaches $v$, then $A$ is a *minimal attack on $v$*. The set of minimal attacks on $v$ is denoted $[\![v]\!]$. We define $f_T(A) \doteq f_T(R_T, A)$, and attacks that reach $R_T$ are called *succesful* w.r.t. $T$. Furthermore, the minimal attacks on $R_T$ (i.e. the minimal successful attacks) are called *minimal attacks*. ATs are *coherent* [46], meaning that adding attack steps preserves success: if $A$ is successful then so is $A \cup \{a\}$ for any $a \in \mathrm{BAS}$. Thus, the suite of successful attacks of an AT is characterised by its minimal attacks.

**Definition 4.** *The* semantics of an AT $T$ *is its suite of minimal attacks $[\![T]\!]$.*

**Example 2.** *Consider the AT in Fig. 3 representing ways to access the ground station database of a CubeSAT as admin: its suite of minimal attacks consists of $\{\{IGP, LDG, LM\}, \{IGP, LDG, EV\}\}$. That is, to mount a minimal attack a malicious actor needs to gain access performing information gathering and phishing IGP — a BAS that is further refined in Fig. 6 — and by logging into the DB of the ground station; to then either leverage misconfigurations LM or exploit vulnerabilities EV in the DB software to gain admin privileges. A non-minimal attack on this AT would include both LM and EV.*

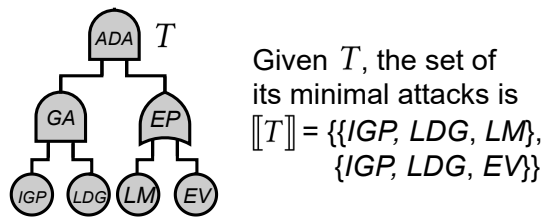

Given $T$, the set of its minimal attacks is $[\![T]\!]$ = {{*IGP, LDG, LM*}, {*IGP, LDG, EV*}}

**Figure 3**: All minimal attacks for AT $T$ modelling access to ground station DB of a CubeSAT (excerpt from Fig. 6).

## 2.1 Security metrics for attack trees

Security metrics — such as the minimal time and cost among all attacks — are essential to perform quantitative analysis of systems and to support more informed decision making processes. To enable this, i.e. computing security metrics, we adopt the well-established *semiring* framework. Semirings have vast applicability potential [47] and have been successfully used to construct attribute domains on ATs [2, 48, 49]. In this paper, we formulate *linearly ordered unital semiring attribute domains* where $V$ is the value domain, $\triangle$ is an operator to combine values of BASes in an attack, $\triangledown$ is an operator to combine values of different attacks and $\preceq$ is an order to compare values. These *linearly ordered unital semiring attribute domains* provide a convenient way to define an ample class of metrics including "min cost", "min time" — both with parallel or sequential attack steps — "min skill" and "discrete probability".

**Definition 5.** *A* linearly ordered unital semiring attribute domain *(simply* attribute domain *or* LOAD *from now on) is a tuple* $L = (V, \triangledown, \triangle, 1_\triangledown, 1_\triangle, \preceq)$ *where:*
- $V$ *is a set;*
- $\triangledown, \triangle \colon V^2 \to V$ *are commutative, associative binary operations on $V$;*
- $\triangle$ *distributes over* $\triangledown$*, i.e.,* $x \triangle (y \triangledown z) = (x \triangledown y) \triangle (x \triangledown z)$ *for all* $x, y, z \in V$;
- $\triangledown$ *is absorbing w.r.t.* $\triangle$*, i.e.,* $x \triangledown (x \triangle y) = x$ *for all* $x, y \in V$;
- $1_\triangledown$ *and* $1_\triangle$ *are unital elements, i.e.,* $1_\triangledown \triangledown x = 1_\triangle \triangle x = x$ *for all* $x \in V$;
- $\preceq$ *is a linear order on $V$.*

As anticipated, many relevant metrics for security analyses on ATs can be formulated as attribute domains. Table 1 shows examples, where $\mathbb{N}_\infty = \mathbb{N} \cup \{\infty\}$ includes 0 and $\infty$.

**Example 3.** *An example of a LOAD is* $(\mathbb{N}_\infty, \min, +, \infty, 0, \leq)$*. Indeed,* min *and* + *are commutative, associative operations on* $\mathbb{N}_\infty$*. The distributive property amounts to the fact that* $x + \min(y, z) = \min(x + y, x + z)$*, while the absorbing property can be stated as* $\min(x, x + y) = x$*. The units are given by* $1_{\min} = \infty$ *and* $1_+ = 0$*, and $\leq$ is a linear order on $\mathbb{N}_\infty$. As we will discuss in Ex. 4,* *this LOAD corresponds to the* min cost *metric on ATs.*

It is important to note that derived metrics such as stochastic analyses and Pareto frontiers can be represented by semirings. However, they do not fit in this framework not being LOADs [2].

| METRIC | $V$ | $\triangledown$ | $\triangle$ | $1_\triangledown$ | $1_\triangle$ | $\preceq$ |
|---|---|---|---|---|---|---|
| min cost | $\mathbb{N}_\infty$ | min | + | $\infty$ | 0 | $\leq$ |
| min time (sequential) | $\mathbb{N}_\infty$ | min | + | $\infty$ | 0 | $\leq$ |
| min time (parallel) | $\mathbb{N}_\infty$ | min | max | $\infty$ | 0 | $\leq$ |
| min skill | $\mathbb{N}_\infty$ | min | max | $\infty$ | 0 | $\leq$ |
| discrete prob. | $[0,1]$ | max | $\cdot$ | 0 | 1 | $\leq$ |

**Table 1**: AT metrics with attribute domains.

Moreover, some meaningful metrics — like the cost to defend against all attacks — do fall outside this category [2]. To render this framework functional, all BASes of ATs are enriched with attributes. More precisely, first an *attribution* $\alpha$ assigns a value to each BAS; then a *security metric* $\widehat{\alpha}$ assigns a value to each attack scenario; and finally the *metric* $\breve{\alpha}$ assigns a value to the set of minimal attacks. We then refer to LOADs to define AT metrics. Given a LOAD $(V, \triangledown, \triangle, 1_\triangledown, 1_\triangle, \preceq)$ we assign to each BAS $a$ an *attribute value* $\alpha(a) \in V$. The operators $\triangledown, \triangle$ are then used to define a metric value for $T$ as follows:

**Definition 6.** *Let $T$ be an AT and let* $L = (V, \triangledown, \triangle, 1_\triangledown, 1_\triangle, \preceq)$ *be a LOAD.*
1. *An* attribution *on $T$ with values in $L$ is a map* $\alpha \colon BAS_T \to V$;
2. *Given such $\alpha$, define the* metric value *of an attack $A$ by*
$$\widehat{\alpha}(A) = \bigtriangleup_{a \in A} \alpha(a);$$
3. *Given such $\alpha$, define the* metric value *of $T$ by*
$$\breve{\alpha}(T) = \bigtriangledown_{A \in [\![T]\!]} \widehat{\alpha}(A) = \bigtriangledown_{A \in [\![T]\!]} \bigtriangleup_{a \in A} \alpha(a).$$

**Example 4.** *Consider $L$ from Ex. 3 representing the metric* min cost*, and let $T$ be the AT in Fig. 4. To each BAS we attach a cost value, given by the attribution* $\alpha \colon BAS_T \to V$ *given by* $\{IGP \mapsto 15, LDG \mapsto 2, LM \mapsto 7, EV \mapsto 9\}$*. As in Ex. 2, $T$ has two minimal attacks, $A_1 = \{IGP, LDG, LM\}$ and $A_2 = \{IGP, LDG, EV\}$. Since $\triangle = +$, We have* $\widehat{\alpha}(A_1) = \alpha(IGP) + \alpha(LDG) + \alpha(LM) = 15 + 2 + 7 = 24$; *this is the cost*

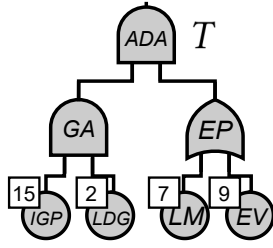

Given $T$ and an attribution over BASes
$\alpha : \{IGP \mapsto 15, LDG \mapsto 2, LM \mapsto 7, EV \mapsto 9\}$,
then *min cost* for $T$ is calculated as follows:

$$\begin{aligned}\check{\alpha}(T) &= \widehat{\alpha}(\{IGP, LDG, LM\}) \triangledown \widehat{\alpha}(\{IGP, LDG, EV\})\\ &= \widehat{\alpha}(\{IGP \triangle LDG \triangle LM\}) \triangledown \widehat{\alpha}(\{IGP \triangle LDG \triangle EV\})\\ &= \min(15+2+7, 15+2+9) = \min(24, 26) = 24.\end{aligned}$$

**Figure 4**: Computing *min cost* for $T$: AT for accessing a ground station DB of a CubeSAT.

an attacker needs to spend to perform attack $A_1$. Similarly one finds $\widehat{\alpha}(A_2) = 15 + 2 + 9 = 26$. We then calculate $\check{\alpha}(T) = \min(\widehat{\alpha}(A_1), \widehat{\alpha}(A_2)) = 24$. Indeed, the minimal cost incurred by an attacker to succesfully attack the system is by performing the cheapest minimal attack, which is $A_1$.

When computing multiple metrics on a given AT, one can resort to multiple LOADs and coherently chosen attributions over its BASes. We thus define such a tree as follows:

**Definition 7.** *An* attributed AT *is a tuple* $\mathsf{T} = (T, \mathscr{L}, \mathfrak{a})$ *where: 1. $T$ is an attack tree; 2. $\mathscr{L} = \{L_1, \ldots, L_l\}$ is a set of LOADs; 3. $\mathfrak{a} = \{\alpha_i\}_{i=1}^{l}$ is a set of attributions on $T$, where each $\alpha_i$ takes values in $L_i$.*

Although in this paper we calculate metrics by considering all *minimal* attacks — coherently with [2] — one could also simply consider all successful attacks. For metrics obtained from LOADs this does not make a difference: for example, the successful attack with minimal cost will always be a minimal attack, since adding BASes can only increase the cost. Therefore, in the calculation of min cost we may as well take the minimum over all successful attacks, rather than just minimal attacks.

## 3 A Logic for AT Metrics

### 3.1 Syntax of ATM

Below, we present ATM, a logic for general Metrics on Attack Trees. ATM shares the objective of developing a language directly on tree-shaped models with [16, 17]. However, it extends the scope of these works to the security domain and allows for property specification that consider a large class of security metrics. The syntax of ATM is structured on four layers. The first layer, $\phi$, reasons about the status of elements in an AT. Atomic formulae $e$ represent BASes and IEs in an AT and they can be combined with usual Boolean connectives. Furthermore, we can forcefully set the value of an element in a layer 1 formula to either 0 or 1 with $\phi[e \mapsto 0]$ and $\phi[e \mapsto 1]$. With $\mathrm{MA}(\phi)$ we can check whether an attack is a *minimal attack*, i.e., a minimal attack successful for a given $\phi$. Layer two and three reason about metrics. Layer 2 formulae allow the user to check whether a given metric on a $\phi$ formula is bounded by $m$ ($\mathbb{M}_k(\phi) \preceq_k m$) and to forcefully set the attribution of a given $e \in \psi$ to an appropriate value $\nu$ ($\psi[e \stackrel{k}{\mapsto} \nu]$). Boolean connectives are also allowed. Layer 3 formulae also allow the setting of attributions but simply return the *value* of a calculated metric ($\mathbb{V}_k(\phi)$). Note that for the layer 1, layer 2 and layer 3 formulae we usually assign values with $\mapsto$ to $e \in \mathsf{BAS}$. We can however assign values to IEs if 1. $e$ is a module [50], i.e., all paths between descendants of $e$ and the rest of the AT pass through $e$ 2. and none of the descendants of $e$ are present in the formula. If so, we prune that (sub-)AT and treat occurring IEs as BASes. Finally, layer 4 formulae allow us to perform quantification over layer 1 and layer 2 formulae. Given a set of LOADs $\mathscr{L} = \{L_1, \ldots, L_l\}$ with $L_k \in \mathscr{L}$ and $m \in V_k$ the syntax is defined as follows:

Layer 1: $\phi ::= e \mid \neg \phi \mid \phi \wedge \phi \mid \phi[e \mapsto 0] \mid \phi[e \mapsto 1]$
$\qquad \mid \mathrm{MA}(\phi)$

Layer 2: $\psi ::= \neg \psi \mid \psi \wedge \psi \mid \mathbb{M}_k(\phi) \preceq_k m \mid \psi[e \stackrel{k}{\mapsto} \nu]$

Layer 3: $\xi ::= \mathbb{V}_k(\phi) \mid \xi[e \stackrel{k}{\mapsto} \nu]$

Layer 4: $\gamma ::= \neg \gamma \mid \exists (\phi \wedge \psi) \mid \forall (\phi \wedge \psi)$

**Syntactic sugar.** We define the following derived operators, where formulae $\theta$ are either layer 1 or layer 2 formulae.

$\theta_1 \vee \theta_2 ::= \neg(\neg\theta_1 \wedge \neg\theta_2)$, $\theta_1 \not\Leftrightarrow \theta_2 ::= \neg(\theta_1 \Leftrightarrow \theta_2)$,
$\theta_1 \Rightarrow \theta_2 ::= \neg(\theta_1 \wedge \neg\theta_2)$, $\quad \text{MD}(\phi) ::= \text{MA}(\neg\phi)$,
$\theta_1 \Leftrightarrow \theta_2 ::= (\theta_1 \Rightarrow \theta_2) \wedge (\theta_2 \Rightarrow \theta_1)$

where $\text{MD}(\phi)$ checks whether $A$ is a *minimal defence* w.r.t. $\phi$, i.e., a set that guarantees that $\phi$ is not reached.

## 3.2 Semantics of ATM

The semantics for our logic reflect objects needed to evaluate the four syntactical layers. For the first layer of ATM, formulae are evaluated on an attack $A$ and on a tree $T$. Atomic formulae $e$ are satisfied by $A$ and $T$ if the structure function in Def. 3 returns 1 with $A$ and $e$ as input. Formally:

$A, T \models e \quad \text{iff } f_T(e, A) = 1$
$A, T \models \neg\phi \quad \text{iff } A, T \not\models \phi$
$A, T \models \phi \wedge \phi' \quad \text{iff } A, T \models \phi \text{ and } A, T \models \phi'$
$A, T \models \phi[e_i \mapsto 0] \text{ iff } A', T \models \phi \text{ with } A' = \{a'_1, \ldots, a'_n\}$
$\qquad\qquad\qquad\qquad\text{where } a'_i = 0 \text{ and } a'_j = a_j \text{ for } j \neq i$
$A, T \models \phi[e_i \mapsto 1] \text{ iff } A', T \models \phi \text{ with } A' = \{a'_1, \ldots, a'_n\}$
$\qquad\qquad\qquad\qquad\text{where } a'_i = 1 \text{ and } a'_j = a_j \text{ for } j \neq i$
$A, T \models \text{MA}(\phi) \quad \text{iff } A \in [\![\phi]\!]_T$

With $[\![\phi]\!]_T$ we denote the *minimal satisfaction set* of attacks for $\phi$, i.e., the set of minimal attacks that satisfy $\phi$ given $T$. We define $[\![\phi]\!]_T$ as follows: $[\![\phi]\!]_T = \{A \mid A, T \models \phi \wedge \not\exists A' \subseteq A. A', T \models \phi\}$. It is important to note that — with semantics defined as we did — we allow for fairly granular reasoning over ATs. In particular, we can evaluate whether an attack compromises a particular sub-AT without *reaching* the TLE. Semantics for the second and third layer require *attributed trees* (see Def. 7). We can then define semantics for the second layer:

$A, \mathsf{T} \models \neg\psi \quad \text{iff } A, \mathsf{T} \not\models \psi$
$A, \mathsf{T} \models \psi \wedge \psi' \quad \text{iff } A, \mathsf{T} \models \psi \text{ and } A, \mathsf{T} \models \psi'$
$A, \mathsf{T} \models \mathbb{M}_k(\phi) \preceq_k m \text{ iff } A, T \models \phi \wedge \widehat{\alpha}(A) \preceq_k m$
$A, \mathsf{T} \models \psi[e_i \stackrel{k}{\mapsto} \nu] \quad \text{iff } A, \mathsf{T}(\mathfrak{a}[\alpha_k(a_i) \stackrel{k}{\mapsto} \nu]) \models \psi$

For an attack $A$ and an attributed tree $\mathsf{T}$ to satisfy $\mathbb{M}_k(\phi) \preceq_k m$, both the attack $A$ and the tree $T$ must satisfy the inner layer 1 formula and the security metric calculated on the attack must respect the given threshold. We let $X_1$ be the set of layer 1 formulae and we define a $\phi$-*security metric* to attribute a value to a layer 1 formula:

**Definition 8.** *A $\phi$-security metric is a function $\breve{\alpha}^\mathsf{T}: X_1 \to V$ defined as follows:*

$$\breve{\alpha}^\mathsf{T}(\phi) = \bigvee_{A \in [\![\phi]\!]_T} \bigwedge_{a \in A} \alpha(a).$$

Note that in Def. 8 some occurrences can lead to the application of the $\breve{\alpha}$ function to the empty set, i.e., when $[\![\phi]\!]_T = \varnothing$. To account for this, we resort to $1_\triangledown$ and $1_\triangle$ for $\triangledown$ and $\triangle$ (see Def. 5). Assuming the case in which $\breve{\alpha}^\mathsf{T}(\phi) \equiv \breve{\alpha}(\varnothing)$, we fix that $\breve{\alpha}^\mathsf{T}(\phi) = 1_\triangledown$; likewise for $\widehat{\alpha}$ and $1_\triangle$. Furthermore, with $\breve{\alpha}^{\mathsf{T}_k}: X_1 \to V_k$ we denote a $\phi$-security metric whose domain and attribution are obtained appropriately from the $k$-est LOAD $L_k \in \mathscr{L}$. We then let $\mathfrak{a}[\alpha_k(a_i) \stackrel{k}{\mapsto} \nu]$ be the attribution on the element $a_i \in A$ via $\alpha_k$ to an arbitrary value $\nu$, chosen appropriately from the domain $V_k$ of $L_k$. Consequently, we define semantics for the third layer. Let $\mathsf{Val}_\mathsf{T}: X_3 \to V_k$ define an evaluation function of layer three formulae in $X_3$:

$\mathsf{Val}_\mathsf{T}(\mathbb{V}_k(\phi)) = \breve{\alpha}^{\mathsf{T}_k}(\phi)$
$\mathsf{Val}_\mathsf{T}(\xi[e_i \stackrel{k}{\mapsto} \nu]) = \mathsf{Val}_{\mathsf{T}(\mathfrak{a}[\alpha_k(a_i) \stackrel{k}{\mapsto} \nu])}(\xi)$

Finally, we can define semantics for the fourth layer containing quantifiers:

$\mathsf{T} \models \neg\gamma \quad \text{iff } \mathsf{T} \not\models \gamma$
$\mathsf{T} \models \exists(\phi \wedge \psi) \text{ iff } \exists A . A, T \models \phi \text{ and } A, \mathsf{T} \models \psi$
$\mathsf{T} \models \forall(\phi \wedge \psi) \text{ iff } \forall A . A, T \models \phi \text{ and } A, \mathsf{T} \models \psi$

## 4 Case Study: Attacking a CubeSAT

CubeSATs are a type of *nanosatellite* typically used for academic and educational purposes [13]: they are usually built in units (or "U") of 10cm x 10cm x 10cm and can be combined to form larger satellites. They are relatively inexpensive to design, build, and launch compared to traditional, larger satellites and they are a popular choice among students, universities, technology pioneers, and crowd-sourced initiatives [14]. To give a sense of the importance of CubeSATs in our orbital ecosystem, we provide a representation of orbiting CubeSATs as of March 2023 in Fig. 5 and an animation in [51]. A total of 153 elements are plotted on the Earth, following data provided by the online database Celestrack [52]. The size of each sphere is exaggerated for visual purposes — a diameter of 500km for each element — and satellites are propagated using the Simplified General Perturbations 4 (SGP4) orbit propagator [53]. As CubeSATs are one of the platforms achieving more consensus in the context of the "New Space" [14, 54], it is fundamental that security risks on these systems are not overlooked. To cater for this need, we showcase how ATM can be applied to specify useful properties on CubeSATs.

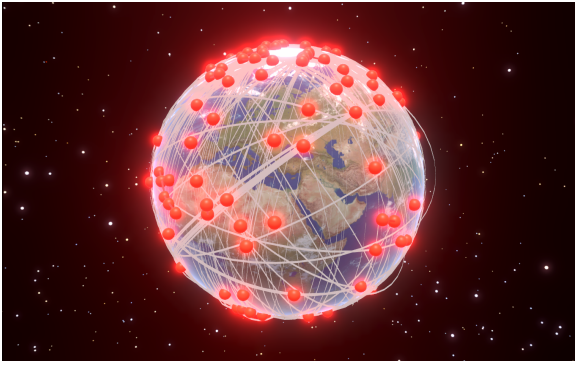

**Figure 5**: Representation of orbiting CubeSATs.

**A CubeSAT AT.** In Fig. 6 (page 10), an AT represents three possible ways in which an attacker could compromise the availability of a CubeSAT. The scenario and the original ATs are taken from [15] and then slightly adapted to model a unique cohesive AT. The TLE in Fig. 6 represents the disruption of CubeSAT's operations — the $DCOp$ OR-gate. This gate is detailed by three children: $DoS$ — the indigo TLE of a sub-tree on the left presenting a denial of service attack — $TDC$ — the violet TLE of the central sub-tree detailing a data tampering attack — and $KR$ — the yellow TLE of the sub-tree on the right that presents an attack killing communications on the CubeSAT. For a denial of service to happen, the attacker must perform information gathering and a successful phishing attack — detailed by the red $IGP$ AND-gate — and use gathered intel to access the CubeSAT UI and disrupt the service. On the other hand, to perform a data tampering attack, one must access the ground control database as admin — detailed by the green $ADA$ AND-gate — then modify database entries and tamper with data. Finally, to kill communications on the CubeSAT an attacker must perform reconnaissance and weaponization, crafting a malicious app, and also conduct the exploit uploading the malware on the CubeSAT via the ground station: executing this code on the satellite would cause communications to go offline. Due to the increasing complexity of these three different attacks, the AT in Fig. 6 presents several sub-trees that are shared. The red sub-tree for information gathering and phishing is shared by the denial of service attack and by the sub-tree that models getting access to the database on the ground station. Furthermore, this green sub-tree is itself shared between the tampering data attack and the more complex malware-based communication killing attack.

**Properties.** ATM allows us to specify some properties on the AT in Fig. 6. As per semantics, properties 2 and 3 are evaluated w.r.t. a given attack. In order to posit some of these queries, we assume that the AT is attributed (see Def. 7).

1) What are all minimal attacks to achieve denial of service?
$$[\![DoS]\!]_T$$

2) Are the cost of data-tampering and info gathering and phishing respectively lower than 20 and at most 5?
$$\mathsf{Cost}(TDC) < 20 \land \mathsf{Cost}(IGP) \leq 5$$

3) Are the probability of successfully attacking the TLE and the parallel time of attack lower than 0.05 and 45 respectively?
$$\mathsf{Prob}(DCOp) < 0.05 \land \mathsf{ParTime}(DCOp) < 45$$

4) What is the min skill an attacker has to have to kill communications on the CubeSAT, assuming that one needs skill of 20 to perform info gathering and phishing?
$$\mathsf{Skill}(KR)[IGP \mapsto 20]$$

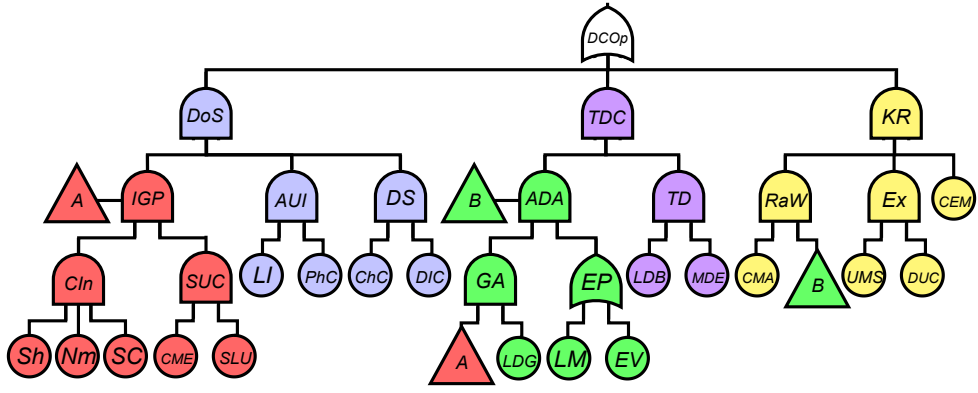

**Figure 6**: An AT representing ways to attack a CubeSAT.

| CubeSAT TLE: | | Info Gathering + Phish: | | Data Tampering: | |
|---|---|---|---|---|---|
| Disrupt CubeSAT Operations | DCOp | Info Gathering & Phishing | IGP | Tamper Data from CubeSAT | TDC |
| **DoS Attack:** | | Collect Information | CIn | Tamper with Data | TD |
| Denial of Service | DoS | Shodan | Sh | Login to DB as Admin | LDB |
| Access CubeSAT UI | AUI | NMAP | Nm | Modify Database Entries | MDE |
| Locate Interfaces | LI | Scrape Credentials | SC | **Kill Comms on CubeSAT:** | |
| Login with Phished Creds | PhC | **Access Ground Station:** | | Kill Radio on CubeSAT | KR |
| Disrupt Service | DS | Access Ground DB as Admin | ADA | Recon. and Weaponization | RaW |
| Change Config. Settings | ChC | Gain Access | GA | Create Malicious App | CMA |
| Delete Items on CubeSAT | DIC | Login to DB Ground Station | LDG | Exploit | Ex |
| Steal User Credentials | SUC | Escalate Privilege | EP | Upload Malware to Server | UMS |
| Craft Malicious Email | CME | Leverage Misconfig. | LM | Command for Upload | EV |
| Send as Legit User | SLU | Exploit Vulnerabilties | EV | SAT Gets & Exec. Malware | CEM |

**Table 2**: Abbreviations for the CubeSAT AT in Fig. 6.

5) Is there an attack that ensures data tampering without exploiting vulnerabilities in the ground station system?

$$\exists (TDC[EV \mapsto 0])$$

6) Is it necessary to leverage misconfigurations to perform a successful attack on CubeSAT's communications?

$$\forall (KR \Rightarrow LM)$$

7) Is there an attack that ensures successful access to the ground station DB while keeping the cost under 20?

$$\exists (\mathsf{Cost}(ADA) < 20)$$

8) Do accessing the CubeSAT UI and disrupting service always imply that successful attacks to the TLE are strictly cheaper than 35 and strictly faster than 60 (when parallelized)?

$$\forall ((AUI \land DS) \Rightarrow (\mathsf{Cost}(DCOp) < 35 \land \mathsf{ParTime}(DCOp) < 60))$$

## 5 Case Study: an AT from MITRE ATT&CK

**The MITRE ATT&CK database.** The MITRE ATT&CK database (*MAd*) is a globally-accessible knowledge base of adversary tactics and techniques based on real-world observations. The ATT&CK knowledge base is used as a foundation for the development of specific threat models and methodologies in the private sector, in government, and in the cybersecurity product and service community [†]. *Campaigns* from *MAd* offer a record of real life attacks via the ATT&CK Navigator: a matrix representation of a given cybercampaign, that organizes recorded (sub)*techniques* by dividing them into *tactics*. (Sub)techniques represent the *how* of a given attack step – e.g., using *Powershell* (technique number T1059.001) – while tactics represent the stage of the attack in which techniques are used, i.e., the *why*: for example, using Powershell in order to perform *Execution* of a malicious task. The *MAd* provides a total of 14 cyberattack tactics: 1. *Reconnaissance*, 2. *Resource Development*, 3. *Initial Access*, 4. *Execution*, 5. *Persistence*, 6. *Privilege Escalation*, 7. *Defence Evasion*, 8. *Credential Access*, 9. *Discovery*, 10. *Lateral Movement*, 11. *Collection*, 12. *Command and Control*, 13. *Exfiltration*, and 14. *Impact* .

---
[†] https://attack.mitre.org/

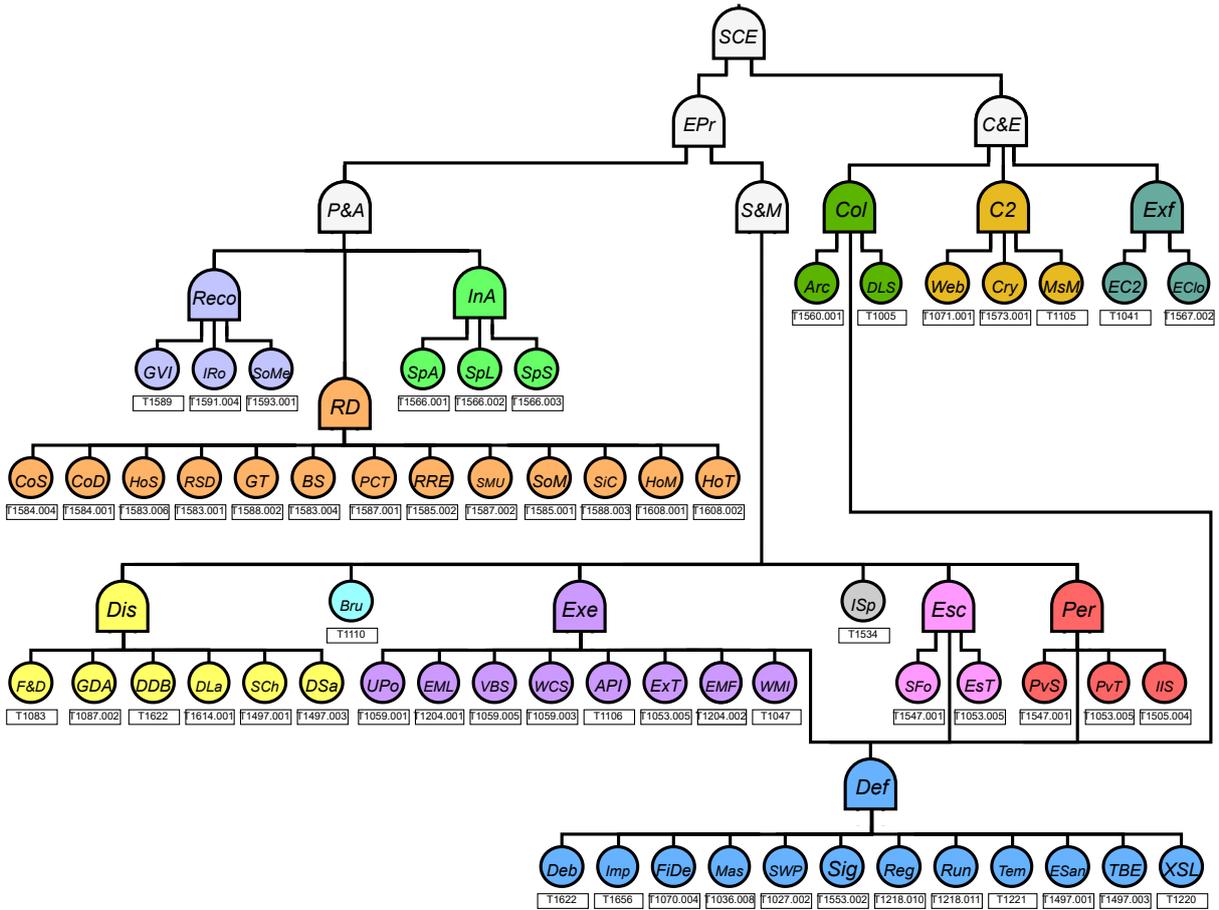

**Figure 7**: An AT modelling *MAd*'s Operation Dream Job (legenda on page 12).

**Operation Dream Job.** We select Operation Dream Job as our candidate real-life cybercampaign affecting the aerospace sector. As reported in the *MAd*, Operation Dream Job[‡] was a cyber espionage operation likely conducted by Lazarus Group[*] that targeted the defense, aerospace, government, and other sectors in the United States, Israel, Australia, Russia, and India.

**Modelling assumptions.** In modelling Operation Dream Job as an AT, we treat every (sub)technique in the ATT&CK Navigator as a BAS: in Fig. 7, a small label for each BAS signals its respective (sub)technique according to the *MAd* nomenclature. Furthermore, we assume that embedding more fine-grained information will be more useful than selecting the coarse-grained equivalent. E.g., if the ATT&CK Navigator for tactic *Reconnaissance* presents both the *Gather Victim Org Information* technique (T1591) and its subtechnique *Identify Roles* (T1591.004) we keep the latter as an AT node and discard the former, due to *Identify Roles* (T1591.004) giving more information on the specifics of this attack step. Finally, we color code nodes (see legenda at page 12) in the AT following *MAd*'s 14 tactics: note that *MAd* does not report any technique affiliated to the *Impact* tactic for this campaign.

---

[‡] https://attack.mitre.org/campaigns/C0022/

[*] https://attack.mitre.org/groups/G0032/

| **Operation Dream Job TLE**: | | | **Discovery**: | | | **Defense Evasion**: | | |
|---|---|---|---|---|---|---|---|---|
| Success in Cyber Espionage | SCE | | Discovery | Dis | | Defense Evasion | Def | |
| Establish Presence | EPr | | Files & Dirs Scan | F&D | | Debugger Evasion | Deb | |
| Collect & Exfiltrate | C&E | | Get Domain Accounts | GDA | | Impersonation | Imp | |
| Prep. & Access | P&A | | Detect Debuggers | DDB | | File Deletion | FiDe | |
| Spread & Maintain | S&M | | Detect System Language | DLa | | File Type Masquerade | Mas | |
| **Reconnaissance**: | | | System Checks | SCh | | Software Packing | SWP | |
| Reconnaissance | Reco | | Detect Sandboxes | DSa | | Code Signing | Sig | |
| Gather Victim Info | GVI | | **Execution**: | | | Use Regsvr32 | Reg | |
| Identify Roles | IRo | | Execution | Exe | | XSL Script Processing | XSL | |
| Social Media | SoMe | | Windows Command Shell | WCS | | Use Rundll32 | Run | |
| **Resource Devel.**: | | | Execute via WMI | WMI | | Template Injection | Tem | |
| Resource Devel. | RD | | VB Malicious Macro | VBS | | Evade Sandboxes | ESan | |
| Get Tools | GT | | Use Powershell | UPo | | Time-Based Evasion | TBE | |
| Buy Server | BS | | Obtain User-Agent via API | API | | **Collection**: | | |
| Prepare Custom Tools | PCT | | Exec with Scheduled Tasks | ExT | | Collection | Col | |
| Register Rogue Email Addr. | RRE | | Exec. via Malicious File | EMF | | Archive via Utility | Arc | |
| Register Same Domain | RSD | | Exec. via Malicious Link | EML | | Data from Local Sys. | DLS | |
| Hosting Services | HoS | | **Credential Access**: | | | **Command & Control**: | | |
| Compromise Domains | CoD | | Brute Force Attacks | Bru | | Command & Control | C2 | |
| Compromise Servers | CoS | | **Lateral Movement**: | | | Web Protocols | Web | |
| Sign Malware & Utils | SMU | | Internal Spearphish. | ISp | | Symmetric Cryptography | Cry | |
| Social Media | SoM | | **Privilege Escalation**: | | | Multistage Malware Ingress | MsM | |
| Signing Certificates | SiC | | Privilege Escalation | Esc | | **Exfiltration**: | | |
| Host Malware | HoM | | Escalate via Startup Folder | SFo | | Exfiltration | Exf | |
| Host Tools | HoT | | Escalate via Task Scheduler | EsT | | Exfil over C2 Channel | EC2 | |
| **Initial Access**: | | | **Persistence**: | | | Exfil over Cloud | EClo | |
| Initial Access | InA | | Persistence | Per | | | | |
| Spearphish. (Attachment) | SpA | | Persist via Startup | PvS | | | | |
| Spearphish. (Link) | SpL | | Persist via Task Scheduler | PvT | | | | |
| Spearphish. (Service) | SpS | | IIS Components | IIS | | | | |

**Table 3**: Abbreviations for the *MAd* AT in Fig. 7.

**Properties.** As seen in Sec. 4 for Fig. 6, we now employ ATM to specify some properties on the AT in Fig. 7. As done in Sec. 4, we assume here that our AT is attributed (see Def. 7).

1) Is there a successful attack for TLE that does not require to *Evade Sandboxes*?

$$\exists (SCE[ESan \mapsto 0])$$

2) Can an attacker always establish presence in the network only by guaranteeing execution of all (sub)techniques for *Resource Development* and *Discovery*?

$$\forall (EPr[RD \mapsto 1, Dis \mapsto 1])$$

3) Is the probability of a successful Operation Dream Job below 0.075, assuming that an attacker can exfiltrate data over cloud with probability 0.015?

$$\mathsf{Prob}(SCE) < 0.075[EClo \mapsto 0.015]$$

4) What is the min (parallel) attack time to *Spread & Maintain* access through the network, assuming that an attacker needs 35 time units for (sub)techniques in *Defense Evasion*?

$$\mathsf{ParTime}(S\&M)[Def \mapsto 35]$$

5) Is there an attack that ensures collection and exfiltration, while keeping the cost under 100?

$$\exists (\mathsf{Cost}(C\&E) < 100)$$

6) Is it sufficient to perform *Defense Evasion* in order to achieve both *Privilege Escalation* and

| Natural Language | Property in ATM | LangATM |
|---|---|---|
| What are all minimal attacks to achieve denial of service? | $[\![DoS]\!]_T$ | **assume:**<br>**computeall:**<br>    MA[DoS] |
| Are the cost of data-tampering and info gathering and phishing respectively lower than 20 and at most 5? | $\mathsf{Cost}(TDC) < 20 \land$ $\mathsf{Cost}(IGP) \leq 5$ | **assume:**<br>**check:**<br>    Cost[TDC] < 20 and<br>    Cost[IGP] $\leq$ 5 |
| Are the probability of successfully attacking the TLE and the parallel time of attack lower than 0.05 and 45 respectively? | $\mathsf{Prob}(DCOp) < 0.05 \land$ $\mathsf{ParTime}(DCOp) < 45$ | **assume:**<br>**check:**<br>    Prob[DCOp] < 0.05 and<br>    ParTime[DCOp] < 45 |
| What is the min skill an attacker has to have to kill communications on the CubeSAT, assuming that one needs skill of 20 to perform info gathering and phishing? | $\mathsf{Skill}(KR)[IGP \mapsto 20]$ | **assume:**<br>    set_skill IGP = 20<br>**compute:**<br>    Skill[KR] |
| Is there an attack that ensures data tampering without exploiting vulnerabilities in the ground station system? | $\exists(TDC[EV \mapsto 0])$ | **assume:**<br>    set EV = 0<br>**check:**<br>    exists TDC |
| Is it necessary to leverage misconfigurations to perform a successful attack on CubeSAT's communications? | $\forall(KR \Rightarrow LM)$ | **assume:**<br>**check:**<br>    forall KR impl LM |
| Is there an attack that ensures successful access to the ground station DB while keeping the cost under 20? | $\exists(\mathsf{Cost}(ADA) < 20)$ | **assume:**<br>**check:**<br>    exists Cost[ADA] < 20 |
| Do accessing the CubeSAT UI and disrupting service always imply that successful attacks to the TLE are strictly cheaper than 35 and strictly faster than 60 (when parallelized)? | $\forall((AUI \land DS) \Rightarrow (\mathsf{Cost}(DCOp) < 35 \land$ $\mathsf{ParTime}(DCOp) < 60))$ | **assume:**<br>    AUI and DS<br>**check:**<br>    forall Cost[DCOp] < 35 and<br>    ParTime[DCOp] < 60 |

**Table 4**: Properties in natural language, ATM and LangATM for the CubeSAT AT in Fig. 6.

*Persistence*?

$$\forall(Def \Rightarrow (Esc \land Per))$$

7) Can an attacker gain access without spearphishing via an email attachment and at the same time ensure that the probability of successfully attacking the TLE is at least 0.01?

$$\exists(InA[SpL \mapsto 0] \land \mathsf{Prob}(SCE) \geq 0.01[SpL \mapsto 0])$$

8) Are the success probabilities for performing the *Execution* and *Escalation* tactics always at least 0.25 and 0.15, if we assume that the probability of successful *Template Injection* is 0.50?

$$\forall((\mathsf{Prob}(Exe) \geq 0.25 \land$$
$$\mathsf{Prob}(Esc) \geq 0.15)[Tem \mapsto 0.50])$$

## 6 LangATM: A DSL for ATM

**Design of LangATM.** To ease usability of our logic, we present LangATM, a Domain Specific Language (DSL) to specify properties in ATM. Defining languages and tools to specify properties and requirements is common: in [55] the authors capture high-level requirements for a steam boiler system in a human readable form with SADL, a controlled English requirements capturing language, and its tool suite ASSERT. Further controlled natural languages for knowledge representation include Processable English (PENG) [56], Controlled English to Logic Translation (CELT) [57] and Computer Processable Language (CPL) [58]. LangATM is constructed by following the same principles of *LangPFL* – a domain specific language for FTs that we developed in previous work [17]. As for *LangPFL*, LangATM is inspired by the PENG, CELT and CPL languages for their ease of use and close proximity to natural language. Furthermore, another notable source of inspiration is FRETish [59], a structured natural language capturing Linear Temporal Logic (LTL). FRETish was developed at NASA and is supported by the FRET tool [60]. Other than for its usability, FRETish inspired our DSL with the distinguishing way in which scope, conditions and components of specified properties are clearly separated from desired behaviours on timing and responses. LangATM expresses only a fragment of ATM: most notably, nesting of formulae is disallowed. By doing so, we retain most of the expressiveness of ATM while making property specification easier. In LangATM, AT elements are referred to with their label and each operator in ATM has a counterpart in the DSL: Boolean operators, not, and, or, impl...; setting the value of AT elements to Boolean or metrics values, set, set_prob, set_cost, set_time, set_skill; MAs and MDs, MA[...], MD[...]; operators to check metrics thresholds/compute metrics values, Prob/Cost/ParTime/SeqTime/Skill[...] $\bowtie$ ..., and Prob/Cost/ParTime/SeqTime/Skill[...] (note that $\bowtie \in \{<, \leq, =, \geq, >\}$); and to perform quantification, i.e. exists/forall.

| Natural Language | Property in ATM | LangATM |
|---|---|---|
| Is there a successful attack for TLE that does not require to *Evade Sandboxes*? | $\exists(SCE[ESan \mapsto 0])$ | **assume:** set ESan = 0 <br> **check:** exists SCE |
| Can an attacker always establish presence in the network only by guaranteeing execution of all (sub)techniques for *Resource Development* and *Discovery*? | $\forall(EPr[RD \mapsto 1, Dis \mapsto 1])$ | **assume:** set RD = 1 <br> set Dis = 1 <br> **check:** forall EPr |
| Is the probability of a successful Operation Dream Job below 0.075, assuming that an attacker can exfiltrate data over cloud with probability 0.015? | $\mathsf{Prob}(SCE) < 0.075[EClo \mapsto 0.015]$ | **assume:** set_prob EClo = 0.015 <br> **check:** Prob[SCE] < 0.075 |
| What is the min (parallel) attack time to *Spread & Maintain* access through the network, assuming that an attacker needs 35 time units for (sub)techniques in *Defense Evasion*? | $\mathsf{ParTime}(S\&M)[Def \mapsto 35]$ | **assume:** set_time Def = 35 <br> **compute:** ParTime[S&M] |
| Is there an attack that ensures collection and exfiltration, while keeping the cost under 100? | $\exists(\mathsf{Cost}(C\&E) < 100)$ | **assume:** <br> **check:** exists Cost[C&E] < 100 |
| Is it sufficient to perform *Defense Evasion* in order to achieve both *Privilege Escalation* and *Persistence*? | $\forall(Def \Rightarrow (Esc \wedge Per))$ | **assume:** Def <br> **check:** forall Esc and Per |
| Can an attacker gain access without spearphishing via an email attachment and at the same time ensure that the probability of successfully attacking the TLE is at least 0.01? | $\exists(InA[SpL \mapsto 0] \wedge \mathsf{Prob}(SCE) \geq 0.01[SpL \mapsto 0])$ | **assume:** set SpL = 0 <br> set_prob SpL = 0 <br> **check:** exists InA and Prob[SCE] $\geq$ 0.01 |
| Are the success probabilities for performing the *Execution* and *Escalation* tactics always at least 0.25 and 0.15, if we assume that the probability of successful *Template Injection* is 0.50? | $\forall((\mathsf{Prob}(Exe) \geq 0.25 \wedge \mathsf{Prob}(Esc) \geq 0.15)[Tem \mapsto 0.50])$ | **assume:** set_prob Tem = 0.50 <br> **check:** forall Prob[Exe] $\geq$ 0.25 and Prob[Esc] $\geq$ 0.15 |

**Table 5**: Properties in natural language, ATM and LangATM for the *MAd* AT in Fig. 7.

**LangATM Templates.** One can specify properties in LangATM by utilizing operators inside structured templates. Assumptions on the status of AT elements can be specified under the **assume** keyword. These assumptions will be automatically integrated in the translated formula accordingly, e.g., set or set_*metric* will be translated with the according operators to set evidence, while other assumptions will be the antecedent of an implication. A second keyword separates specified formulae from the assumptions and dictates the desired result: **compute** and **computeall** compute and return desired values, i.e., metrics values and lists of MAs/MDs respectively, while **check** establishes if a specified property holds.

**Case studies.** In Table 4 and Table 5 we showcase ATM properties specified in Sec. 4 and Sec. 5 with their respective translation in LangATM.

# 7 Model Checking Algorithms

In this section we present model checking algorithms for ATM. As noted in [16, 17], some scenarios, especially in the Boolean domain, are trivial: e.g., checking if $A, T \models \phi$ holds is trivial if $\phi$ is a formula that does not contain a MA or MD operator. In that case, we can simply substitute the values of $A$ in the atoms of $\phi$ and see if the Boolean expression evaluates to true. Non trivial scenarios arise if $\phi$ contains a MA or MD operator or if ATs are not tree-shaped. These require computations based on BDDs, introduced in Sec. 7.1: a coherent choice with the landscape of algorithms for FT logics [16, 17] and AT computation [2]. In this section we build upon these results and present algorithms to: 1) Obtain BDDs from layer 1 formulae taking the structure of a given tree $T$ into account (Sec. 7.2); 2) a) Check whether an attack $A$ and a tree $T$ satisfy a layer 1 formula and b) compute all the satisfying attacks $A$ for a given tree $T$ and layer 1 formula (Sec. 7.3); 3) Check whether an attack $A$ and an attributed tree T satisfy a layer 2 formula (Sec. 7.4); 4) Compute the metric value of a given layer 3 formula (Sec. 7.5); 5) Check whether an attributed tree T satisfies a layer 4 formula (Sec. 7.6).

## 7.1 Binary Decision Diagrams (BDDs)

BDDs are directed acyclic graphs (DAGs) that compactly represent Boolean functions [61] by reducing redundancy. Depending on variable's ordering, BDD's size can grow linearly in the number of variables and at worst exponentially. In practice, BDDs are heavily used, including in AT analysis [2] and in their safety counterpart, FTs [16, 17, 62, 63]. Formally, a BDD is a rooted DAG $B_f$ that represents a Boolean function $f \colon \mathbb{B}^n \to \mathbb{B}$ over variables $Vars = \{x_i\}_{i=1}^n$. Each nonleaf $w$ has two outgoing arrows, labeled 0 and 1, and a label $Lab(w) \in Vars$; furthermore, each leaf has a label 0 or 1. Given a $b$ in $\mathbb{B}^n$, the BDD is used to compute $f(b)$ as follows: starting from the top, upon arriving at a node w with $Lab(w) = x_i$, one takes the 0-edge if $b_i = 0$ and the 1-edge if $b_i = 1$. The label of the leaf one ends up in, is then equal to $f(b)$. A function $f$ can be represented by multiple BDDs, but has a unique *reduced ordered* representative, or ROBDD [64, 65], where the $x_i$ occur in ascending order, and the BDD is reduced as much as possible by removing irrelevant nodes and merging duplicates. This is formally defined below; we let $Low(w)$ (resp. $High(w)$) be the endpoint of $w$'s 0-edge (resp. 1-edge) and let $R_B$ be the BDD root.

**Definition 9.** *Let Vars be a set. A* (RO)BDD *over Vars is a tuple* $B = (W, H, Lab, u)$ *where* $(W, H)$ *is a rooted directed acyclic graph, and* $Lab \colon W \to Vars \sqcup \{0, 1\}, u \colon H \to \{0, 1\}$ *are maps such that:*
1. *Every nonleaf $w$ has exactly two outgoing edges $h, h'$ with $u(h) \neq u(h')$, and $Lab(w) \in Vars$;*
2. *Every leaf $w$ has $Lab(w) \in \{0, 1\}$.*
3. *Vars are equipped with a total order, $B_f$ is thus defined over a pair $\langle Vars, < \rangle$;*
4. *the variable of a node is of lower order than its children, that is: $\forall w \in W_n. Lab(w) < Lab(Low(w)), Lab(High(w))$;*
5. *the children of nonleaf nodes are distinct nodes;*
6. *nodes are uniquely determined by their label, low child and high child.*

## 7.2 BDDs from ATs and layer 1 formulae

The first step to enable further computations is to obtain BDDs from layer 1 formulae taking the structure of a given tree $T$ into account (for related procedures on FT logics see [16, 17]). Following, operations between BDDs are represented by **bold** operands e.g., $\bm{\wedge}, \bm{\vee}$. Where convenient notationally, we write $B_T^\phi$ for $B_T(\phi)$, i.e., the BDD $B$ of $\phi$, given $T$. Given a set of variables $Vars = \{x_i\}_{i=1}^n$ existential quantification can be defined as follows: $\bm{\exists} x. B = \text{RESTRICT}(B, x, 0) \bm{\vee} \text{RESTRICT}(B, x, 1)$ and $\bm{\exists} Vars. B = \bm{\exists} x_1. \bm{\exists} x_2. \ldots \bm{\exists} x_n. B$. Furthermore, we define a set of primed variables $Vars' = \{x_i'\}_{i=1}^n$ and let $B_T^\phi[Vars \leadsto Vars']$ be the BDD $B_T^\phi$ in which every variable $x_i \in Vars$ is renamed to its primed $x_i' \in Vars'$. Finally we let $Vars' \subset Vars \equiv (\bm{\bigwedge}_i x_i' \Rightarrow x_i) \bm{\wedge} (\bm{\bigvee}_i x_i' \neq x_i)$. Algorithms to conduct typical BDD

operations — such as RESTRICT — can be found in [61, 64].

**Definition 10.** *The translation function of an AT $T$ is a function $\mathfrak{f}_T$: E $\to$ BDD that takes as input an element $e \in$ E.*

$$\mathfrak{f}_T(e) = \begin{cases} B(e) & \text{if } e \in \text{BAS} \\ \bigvee_{e' \in ch(e)} \mathfrak{f}_T(e') & \text{if } e \in \text{IE} \text{ and } t(e) = \text{OR} \\ \bigwedge_{e' \in ch(e)} \mathfrak{f}_T(e') & \text{if } e \in \text{IE} \text{ and } t(e) = \text{AND} \end{cases}$$

*where $B(e)$ is a BDD with a single node $w$ with $Low(w) = \mathtt{0}$ and $High(w) = \mathtt{1}$.*

Algo. 1 computes $B_T^\phi$ following semantics for layer 1 formulae:

---
**Algorithm 1** Compute $B_T^\phi$ from $T$ and $\phi$
---
1: **Input**: AT $T$, formula $\phi$
2: **Output:** BDD $B_T^\phi$
3: **Method:**
4: **if** $\phi = e$ **then return** $\mathfrak{f}_T(e)$
5: **else if** $\phi = \neg \phi'$ **then**
6:     **return** $\neg$ (*Algo.* 1($T, \phi'$))
7: **else if** $\phi = \phi' \wedge \phi''$ **then**
8:     **return** *Algo.* 1($T, \phi'$)$\wedge$*Algo.* 1($T, \phi''$)
9: **else if** $\phi = \phi'[e_i \mapsto 0]$ **then**
10:     **return** RESTRICT(*Algo.* 1($T, \phi'$), $x_i, 0$)
11: **else if** $\phi = \phi'[e_i \mapsto 1]$ **then**
12:     **return** RESTRICT(*Algo.* 1($T, \phi'$), $x_i, 1$)
13: **else**     // $\phi = \text{MA}(\phi')$
14:     **return** *Algo.* 1($T, \phi'$)$\wedge$($\neg \exists$ *Vars'*.(*Vars'* $\subset$ *Vars*)$\wedge$*Algo.* 1($T, \phi'$)[*Vars* $\rightsquigarrow$ *Vars'*])
15: **end if**

### 7.3 Model checking layer 1 formulae

**Is an attack successful w.r.t. $\phi$?.** Algo. 2 checks whether $A, T \models \phi$, given an attack $A$ and a tree $T$. First, the BDD $B_T^\phi$ for $\phi$ given $T$ is constructed via Algo. 1. Then, the algorithm walks the BDD path representing values of BASes in $A$. If it ends up in the terminal 0, then $A, T \not\models \phi$, otherwise — if the terminal node is 1 — $A, T \models \phi$.

---
**Algorithm 2** Check if $A, T \models \phi$
---
1: **Input**: attack $A$, attack tree $T$, formula $\phi$
2: **Output:** *true* iff $A, T \models \phi$; *false* otherwise.
3: **Method:**
4: $B_T^\phi \leftarrow$ *Algo.* 1($T, \phi$); $w_i = R_{B_T^\phi}$
5: **while** $w_i \notin W_t$ **do**:
6:     **if** $a_i \in A = 0$ **then** $w_i = Low(w_i)$
7:     **else if** $a_i \in A = 1$ **then** $w_i = High(w_i)$
8:     **end if**
9: **end while**
10: **if** $Lab(w_i) = \mathtt{0}$ **then return** *false*
11: **else**    // $Lab(w_i) = \mathtt{1}$
12:     **return** *true*
13: **end if**

---

**All successful attacks w.r.t. $\phi$..** Our ability to construct a BDD $B_T^\phi$ for layer 1 formulae granted by Algo. 1 allows us to compute all attacks $A$ such that $A, T \models \phi$. Algo. 3 performs this computation by applying the ALLSAT [61] algorithm to $B_T^\phi$: ALLSAT walks down the BDD and stores the paths that lead to the terminal node 1. These paths then represent satisfying attacks for $\phi$ given $T$. Note that Algo. 3 can be used to compute all the minimal attacks of a given $\phi$ by simply calling it on MA($\phi$).

---
**Algorithm 3** Compute all $A$ s.t. $A, T \models \phi$
---
1: **Input**: AT $T$, formula $\phi$
2: **Output:** $\{A \mid A, T \models \phi\}$
3: **Method:**
4: $B_T^\phi \leftarrow$ *Algo.* 1($T, \phi$); $w_i = R_{B_T^\phi}$
5: $\{A \mid A, T \models \phi\} \leftarrow$ ALLSAT($w_i$)
6: **return** $\{A \mid A, T \models \phi\}$

### 7.4 Model checking layer 2 formulae

Algo. 4 presented in this subsection checks if an layer 2 formula is satisfied, given an attack $A$ and an attributed tree T. Boolean connectives are resolved as usual via case distinction. To check whether $A, \mathsf{T} \models \mathbb{M}_k(\phi) \preceq_k m$, first the BDD $B_T^\phi$ for the inner layer 1 formula is constructed and Algo. 2 is emplyed to assess whether $A, T \models \phi$. If that is not the case, we return *false*. Otherwise, we compute the metric value for the given attack following the interpretation of $\triangle$ taken from the $k-$est LOSG $L_k$ of our attributed tree T. We store this value in `metr_val`, and we return the result of the comparison with $\preceq_k m$. To handle the case in which we set evidence for a specific atom $e_i$ in a layer 2 formula, we simply call the algorithm again and we make sure that the attribution $\alpha_k$ of the corresponding $a_i$ is mapped to the chosen value $\nu$.

---

**Algorithm 4** Check if $A, \mathsf{T} \models \psi$

---

1: **Input**: attack $A$, attributed AT $\mathsf{T}$, formula $\psi$
2: **Output**: *true* iff $A, \mathsf{T} \models \psi$; *false* otherwise.
3: **Method:**
4: **if** $\psi = \neg \psi'$ **then return not** *Algo.* 4$(A, \mathsf{T}, \psi')$
5: **else if** $\psi = \psi' \wedge \psi''$ **then return** *Algo.* 4$(A, \mathsf{T}, \psi')$ **and** *Algo.* 4$(A, \mathsf{T}, \psi'')$
6: **else if** $\psi = \mathbb{M}_k(\phi) \preceq_k m$ **then**
7:     **if** *Algo.* 2$(A, T, \phi)$ returns *true* **then** // $A, T \models \phi$
8:        $\mathtt{metr\_val} = \bigtriangleup_{k} \atop a \in A \alpha_k(a)$
9:        **return** $\mathtt{metr\_val} \preceq_k m$
10:     **else** // $A, T \not\models \phi$
11:        **return** *false*
12:     **end if**
13: **else** // $\psi = \psi'[e_i \xmapsto{k} \nu]$
14:     **return** *Algo.* 4$(A, \mathsf{T}(\mathfrak{a}[\alpha_k(a_i) \xmapsto{k} \nu]), \psi')$
15: **end if**

---

### 7.5 Compute metrics for layer 3 formulae

This subsection showcases an algorithm to compute a metric value for a specified $\xi$-formula. If $\xi$ equals $\mathbb{V}_k(\phi)$, one approach would be to directly use the formula of Def. 8. However, directly finding all minimal attacks on $\phi$ is computationally expensive [2]. Instead, we calculate metrics by applying the BDD-based method from [2]. This method exploits the fact that paths from the root to $\mathtt{1}$ in a BDD encode succesful attacks, and $\mathtt{1}$-labeled edges on such a path represent the BAS of these attacks. Assigning weight $\alpha(Lab(w))$ to an edge $(w, High(w))$, the metric value can then be computed by a variant of the shortest path algorithm for DAGs. Note that the method in [2] is defined only for $\phi = e$, but the result readily generalizes. If $\xi = \xi'[e_i \xmapsto{k} \nu]$, the algorithm is called again on $\xi'$ and the attribution $\alpha_k$ on the corresponding $a_i$ is set to $\nu$.

---

**Algorithm 5** Compute metric for $\xi$-formula

---

1: **Input**: attributed AT $\mathsf{T}$, formula $\xi$
2: **Output**: metric value for $\xi$.
3: **Method:**
4: **if** $\xi = \mathbb{V}_k(\phi)$ **then**
5:     $(W, H, Lab, u) \leftarrow$ *Algo.* 1$(T, \phi)$
6:     $W_{\mathrm{todo}} \leftarrow W$
7:     **while** $W_{\mathrm{todo}} \neq \varnothing$ **do**
8:        Take $w \in W_{\mathrm{todo}}$ without children in $W_{\mathrm{todo}}$
9:        **if** $Lab(w) = \mathtt{0}$ **then** $v(w) \leftarrow 1_{\triangledown}$
10:        **else if** $Lab(w) = \mathtt{1}$ **then** $v(w) \leftarrow 1_{\triangle}$
11:        **else**
12:           $v(w) \leftarrow v(Low(w)) \triangledown (v(High(w)) \triangle$
13:           $\alpha(Lab(w)))$
14:        **end if**
15:        $W_{\mathrm{todo}} \leftarrow W_{\mathrm{todo}} \setminus \{w\}$
16:     **end while**
17:     **return** $v(\mathrm{R}_{W,H,Lab,u})$
18: **else** // $\xi = \xi'[e_i \xmapsto{k} \nu]$
19:     **return** *Algo.* 5$(\mathsf{T}(\mathfrak{a}[\alpha_k(a_i) \xmapsto{k} \nu]), \xi')$
20: **end if**

---

### 7.6 Model checking layer 4 formulae

We present an algorithm to check whether an attributed tree $\mathsf{T}$ satisfies a layer 4 formula. The non-trivial cases of Algo. 6 check whether $\mathsf{T} \models \exists (\phi \wedge \psi)$ and $\mathsf{T} \models \forall (\phi \wedge \psi)$. In the former case, for each attack $A_i$ in the set of satisfying attacks for $\phi$ — $\{A \mid A, T \models \phi\} \leftarrow$ *Algo.* 3$(T, \phi)$ — we check whether $A_i, \mathsf{T} \models \psi$. If we find a fitting $A_i$, we return it alongside *true*. Otherwise, we return *false*. In the latter case, for each $A_i$ in the set of all attacks for $T$ $\mathscr{A}_T$ we check whether either $A_i, T \not\models \phi$ or $A_i, \mathsf{T} \not\models \psi$. If we find a counterexample $A_i$, we return it alongside *false*. Otherwise, we return *true*.

---

**Algorithm 6** Check if $\mathsf{T} \models \gamma$

---

1: **Input**: set of all attacks $\mathscr{A}_T$, attributed AT $\mathsf{T}$, formula $\gamma$
2: **Output**: *true* iff $\mathsf{T} \models \gamma$; *false* otherwise; (counter)example $A_i$.
3: **Method:**
4: **if** $\gamma = \neg \gamma'$ **then return not** *Algo.* 6$(A, \mathsf{T}, \gamma')$
5: **else if** $\gamma = \exists (\phi \wedge \psi)$ **then**
6:     **for** $A_i \in \{A \mid A, T \models \phi\} \leftarrow$ *Algo.* 3$(T, \phi)$ **do**
7:        **if** *Algo.* 4$(A_i, \mathsf{T}, \psi)$ returns *true* **then return** *true*, $A_i$
8:        **end if**
9:     **end for**
10:     **return** *false*
11: **else if** $\gamma = \forall (\phi \wedge \psi)$ **then**
12:     **for** $A_i \in \mathscr{A}_T$ **do**
13:        **if** *Algo.* 2$(A_i, T, \phi)$ returns *false* $\vee$ *Algo.* 4$(A_i, \mathsf{T}, \psi)$ returns *false* **then**

```
14:            return false, A_i
15:        end if
16:    end for
17:    return true
18: end if
```

## 8 Conclusions

We presented ATM, a logic for general metrics on ATs that enables the construction of complex queries and insightful what-if scenarios. We showcased its usefulness with an application of ATM to the case study of a CubeSAT and to the model of a real-life cyberespionage campaign as recorded by the MITRE ATT&CK Database. Furthermore, we presented LangATM – a domain specific language to ease property specification in ATM. Specified properties can then be checked and metrics computed via model checking algorithms that we provided. Our work opens several relevant perspectives for future research. First, it would be interesting to extend ATM to consider timed behaviours: this would allow to further extend quantitative analysis capabilities. This step could be achieved by extending ATM to dynamic ATs that consider the sequential nature of attack steps. To handle dynamic gates in dynamic ATs it would be very natural to have a logic that can express temporal properties, moving more in the direction of LTL [20] or CTL [19] or their timed variants TLTL [66] and TCTL [67]. Another notable extension of ATM could express and calculate Pareto fronts between metrics [2]. Moreover, it is foreseeable to extend the proposed framework to safety-security variants of ATs and FTs, e.g., to attack-fault trees (AFTs) [68], and to graphs that consider more general safety-security *risks*, in the sense of *probability × impact* [69]. Lastly, implementing this logic could further propel usability of ATM by providing hands-on feedback from domain experts acquainted with threat modelling and vulnerability analysis.

**Acknowledgements.** The authors would like to thank Dr. Juan A. Fraire  (Inria, CONICET and Saarland University) for the insightful discussions about routing in space and for propagating and visualizing orbiting CubeSATs, resulting in Fig. 5 and in the animation in [51].

## Declarations

This work was partially funded by the European Union's Horizon 2020 research and innovation programme under the Marie Skłodowska-Curie grant agreement No 101008233, and the ERC Consolidator Grant 864075 (*CAESAR*).